\documentclass[preprint]{aastex}

\begin{document}

\title{The 4th IBIS/ISGRI soft gamma-ray survey
catalog\footnote{\rm Based on observations with INTEGRAL, an ESA project
with instruments and science data centre funded by ESA member states
(especially the PI countries: Denmark, France, Germany, Italy,
Switzerland, Spain), Czech Republic and Poland, and with the
participation of Russia and the USA.}}

\author{A. J. Bird\altaffilmark{1},
    A. Bazzano\altaffilmark{2},
    L. Bassani\altaffilmark{3},
    F. Capitanio\altaffilmark{2},
    M. Fiocchi\altaffilmark{2},
    A. B. Hill\altaffilmark{1,4},
    A. Malizia\altaffilmark{3},
    V. A. McBride\altaffilmark{1},
    S. Scaringi\altaffilmark{1},
    V. Sguera\altaffilmark{3},
    J. B. Stephen\altaffilmark{3},
    P. Ubertini\altaffilmark{2},
    A. J. Dean\altaffilmark{1},
    F. Lebrun\altaffilmark{5,6},
    R. Terrier\altaffilmark{5},
    M. Renaud\altaffilmark{5},
    F. Mattana\altaffilmark{5},
    D. Gotz\altaffilmark{6},
    J. Rodriguez\altaffilmark{6},
    G. Belanger\altaffilmark{7,5},
    R. Walter\altaffilmark{8},
    C. Winkler\altaffilmark{9}
    }

\altaffiltext{1}{School of Physics and Astronomy, University of Southampton, SO17 1BJ, UK}
\altaffiltext{2}{IASF/INAF, Rome, Italy}
\altaffiltext{3}{IASF/INAF, Bologna, Italy}
\altaffiltext{4}{Laboratoire d'Astrophysique de Grenoble, UMR 5571 CNRS, Universit\'{e} Joseph Fourier, BP 53, 38041 Grenoble, France}
\altaffiltext{5}{AstroParticule et Cosmologie (APC), CNRS-UMR 7164, Universit\'{e} Paris VII, Paris, France}
\altaffiltext{6}{CEA Saclay, DSM/Irfu/Service d'Astrophysique, F-91191, Gif-sur-Yvette, France}
\altaffiltext{7}{ESA/ESAC, PO Box 78, 28691 Villanueva de la Canada, Spain}
\altaffiltext{8}{ISDC, Geneva Observatory, University of Geneva, Chemin d'Ecogia 16, 1291 Versoix, Switzerland}
\altaffiltext{9}{ESA-ESTEC, Research and Scientific Support Dept., Keplerlaan 1, 2201 AZ, Noordwijk, The Netherlands}


\begin{abstract}

In this paper we report on the fourth soft gamma-ray source catalog obtained with the IBIS gamma-ray imager on board the {\em INTEGRAL} satellite. The scientific dataset is based on more than 70~Ms of high quality observations performed during the first five and a half years of Core Program and public observations. Compared to previous IBIS surveys, this catalog includes a substantially increased coverage of extragalactic fields, and comprises more than 700 high-energy sources detected in the energy range 17--100 keV, including both transients and faint persistent objects which can only be revealed with longer exposure times. A comparison is provided with the latest {\em Swift}/BAT survey results.

\end{abstract}



\section{Introduction}

Since its launch in 2002, the {\em INTEGRAL} (International Gamma-Ray Astrophysics Laboratory) observatory has carried out more than 7 years of observations in the energy range from 5 keV -- 10 MeV. {\em INTEGRAL} is an observatory-type mission, and most of the total observing time (65\% in the nominal phase, 75\% during the mission extension) is awarded as the General Programme to the scientific community at large. Typical observations last from 100 ks up to two weeks. As a return to the international scientific collaborations and individual scientists who contributed to the development, design and procurement of {\em INTEGRAL}, a part of the observing time (from 35\% to 25\%) was allocated to the Core Programme. During the nominal lifetime (5 years) this programme consisted of three elements, a deep exposure of the Galactic central radian, regular scans of the Galactic Plane, pointed observations of the Vela region and Target of Opportunity follow-up. In order to exploit {\em INTEGRAL}'s unique capabilities, {\em Key Programmes} were introduced in 2006 (AO5). These are deep observations requesting a few Ms observing time that allow the observatory to accommodate various different requests of the community at large by amalgamating many individual scientific targets present in the selected sky fields as well as ultra-long nucleosynthesis and diffuse emission studies.

The IBIS (Imager on Board {\em INTEGRAL} spacecraft) imaging instrument is optimised for survey work with a large ($30^{\circ}$) field of view with excellent imaging and spectroscopy capability. Instrumental details and sensitivity can be found in \citet{Ubertini2003}. The data are collected with the low-energy array, ISGRI ({\em INTEGRAL} Soft Gamma-Ray Imager; \cite{Lebrun2003}), consisting of a pixellated 128x128 CdTe solid-state detector that views the sky through a coded aperture mask. IBIS/ISGRI generates images of the sky with a 12$'$ (FWHM) resolution and typical source location of better than 1$'$ over a $\sim 19^{\circ}$(FWHM) field of view in the energy range 17--1000 keV.

A sequence of IBIS survey catalogs have been published at regular intervals as more data have become available (Table~\ref{table:surveys}). The frequent Galactic Plane Scans (GPS) within the Core Programme, performed in the first year of operations, were successfully exploited to yield a first survey of the galactic plane to a depth of $\sim$1mCrab in the central radian \citep{Bird2004cat1}. This gave evidence of a soft gamma-ray sky populated with more than 120 sources, including a substantial fraction of previously unseen sources. The second IBIS/ISGRI catalog \citep{Bird2006cat2} used a greatly increased dataset (of $\sim$10Ms) to unveil a soft gamma-ray sky comprising 209 sources, again with a substantial component ($\sim$25\%) of new and unidentified sources. The third IBIS/ISGRI catalog \citep{Bird2007cat3} further increased the dataset, with a substantial improvement in extragalactic coverage, resulting in the detection of a total of 421 sources.

In this paper we provide the fourth IBIS/ISGRI soft gamma-ray survey catalog, that now comprises more than 700 high-energy sources. This fourth catalog continues to build on the source data provided by previous catalogs by incorporating an additional 2 years of data, and using the latest software and source detection techniques. Particular care has been taken to optimise the detection of the transient sources that are common in the hard X-ray sky but are only visible for a small fraction of the total exposure now available.

\begin{table}[htbp]
\centering
\caption{Summary of the IBIS survey catalogs so far \label{table:surveys} }
\begin{tabular}{|c|c|c|c|} \hline
Cat & Exposure & Dates & Sources \\ \hline
1 & 5 Ms & Feb 03 -- Oct 03 & 120\\
2 & 10 Ms & Feb 03 -- June 04 & 209\\
3 & 40 Ms & Feb 03 -- Apr 06 & 421\\
4 & 70 Ms & Feb 03 -- Apr 08 & 723\\ \hline
\end{tabular}
\end{table}


\section{Data analysis and catalog construction}

\subsection{Input dataset and pipeline processing \label{pipeline}}

The survey input dataset consists of all available pointings at the end of April 2008. This consists of the first 5 years of Core Programme observations, including the Galactic Plane Scans (GPS), Galactic Center Deep Exposure (GCDE) and all available pointed observations. Data coverage from revolution 12 (first light, November 2002) to revolution 530 (April 2007) is almost complete, while data between April 2007 and April 2008 constitute only Core Programme and public pointings. {\em INTEGRAL}/IBIS data is organised in short pointings (science windows, scw) of $\sim$2000s. In total, 41588 science windows were input into the pipeline processing. After removal of pointings flagged as Bad Time Intervals (BTI) by the Science Data Centre \citep{Courvoisier2003} this number is reduced to 39548 science windows of good quality data.  

Pipeline processing was carried out using the standard {\em OSA 7.0} software \citep{Goldwurm2003} up to and including the production of sky images for individual science windows with 4.8$'$ pixels. Five primary energy bands (20--40, 30--60, 20--100, 17--30, 18--60 keV) were used to maximise detection sensitivity for sources with various energy spectra. The input catalog used for image processing was those sources marked as detected by ISGRI in the ISDC Reference Catalog version 28, which had been updated to include all sources previously detected by the surveys and in guest observer pointings.

\begin{figure}[htbp]
\centering
\includegraphics[width=0.75\columnwidth]{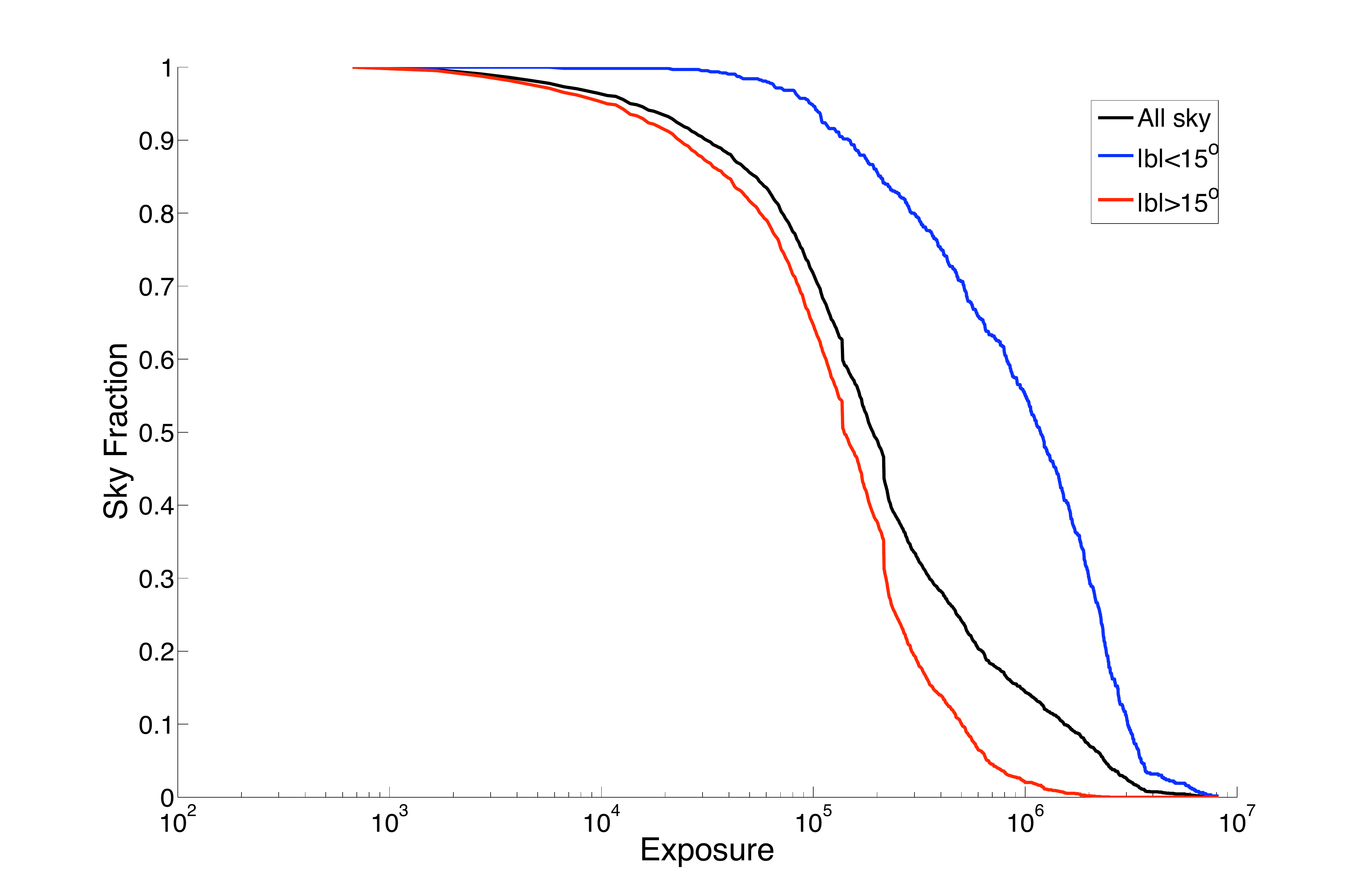}
\vspace{0.5cm}
\caption{Fractional exposure as function of sky area.
\label{fig:expofrac}}
\end{figure}

The overall sky exposure is summarised in Figure~\ref{fig:expofrac}. When discussing exposure, we use the accumulated instrument livetime, corrected for off-axis coding fraction, but not corrected for energy-dependent on-axis absorption. It can be seen that near the Galactic plane, half the sky is covered with more than 1Ms of exposure, while for the whole sky, that fraction drops to $\sim$15\%. 90\% of the sky is exposed at the 100ks or greater level. The exposure does not result from any specific pointing or operational constraints, but is merely the summation of all science observations performed during the accumulation of the dataset. The overall exposure uniformity is improving as the mission continues, and as the science program includes a greater number and diversity of targets.

\subsection{Mosaic construction}

Each scw image was tagged with its rms (after removal of sources) to act as an indicator of overall image quality. As in previous survey construction, the primary aim of this step is to remove data taken during periods of enhanced background (during solar activity or soon after perigee passage). Filtering was applied based on the rms value of the image, such that the rms should not exceed a limit of 2$\sigma$ above the mean image rms for the whole dataset). This function now somewhat overlaps with the BTI flagging provided by the ISDC, but we still remove around 5\% of the science windows that exceed this rms limit. Although they are still processed, data taken in staring mode are not used in the construction of the final sky mosaic images as they contribute a far higher level of systematic noise than the standard dithered observations. Some 1230 science windows in the input dataset were flagged as consisting of staring data.

After removal of high-rms and staring data, approximately 36000 scw remained in the dataset, with a total exposure of $\sim$70Ms. The selected science windows were mosaicked using a proprietary tool optimised to create all-sky galactic maps based on large numbers of  input science windows. Mosaics were constructed for five energy bands (see Section \ref{pipeline}), four map projections, and four timescales, all with 2.4$'$ pixel resolution, significantly oversampling the intrinsic system PSF.

We constructed mosaics over the same three timescales used for the third catalog, and a newly introduced one for identifying transients as explained in section~\ref{bursticity}. Maps were created for each revolution (a satellite orbit; approximately 3 days) which contained valid data. This is optimised to detect sources active on timescales of the order of a day.  We also identified 32 sequences of consecutive revolutions which had similar pointings. Thus these {\em revolution sequences} could best be analysed as a single observation, and sensitivity for sources on longer timescales than revolutions (i.e. order of weeks) could be optimised. Ultimately, persistent sources can best be detected in an {\em all-archive} accumulation of all available high-quality data. The problem of higher exposure and long timebase spanned by this latest dataset has further worsened the problem noted in the third catalog -  namely that the source search methods we employ are optimised for detection of persistent flux from a source; a highly variable source may be clearly detectable during outburst, while having an undetectably low mean flux over the full dataset. In addition, we searched for the optimum detection timescale (from 0.5 days to the length of the dataset) for known or suspected sources (see section~\ref{bursticity}) and created one additional mosaic for each source on the optimum timescale for that source.

For each energy band and time period, all-sky mosaics were made in four projections: centred on the Galactic Center, centred on the Galactic anti-center, north galactic polar and south galactic polar. The purpose of these multiple projections is to present the automatic source detection algorithms with source PSFs with the minimum possible distortions.

\subsection{Source searching and candidate list production}

In total over 11500 maps were created at this stage of processing. Each of the mosaics was searched using two methods: 

\begin{itemize}

\item[(i)] the {\it SExtractor 2.5} software \citep{sextractor}. The source positions measured by {\it SExtractor} represent the centroid of the source calculated by taking the first order moments of the source profile (referred to by {\it SExtractor} as the barycenter method). Source detectability is limited at the faintest levels by background noise and can be improved by the application of a linear filtering of the data.  In addition, source confusion in crowded fields can be minimised by the application of a bandpass filter.  To this end, the {\it mexhat} bandpass filter is used in the {\it SExtractor} software. The convolution of the filter with the mosaic alters the source significances, hence {\it SExtractor} uses the source positions identified from the filtered mosaic to extract the source significances from the original mosaic.

\item[(ii)] a proprietary `peakfind' tool which employs a basic iterative removal of sources technique, combined with an assessment of the local background rms to reduce the false detection of sources in areas of the map with high systematic noise structures - mainly in crowded regions and around the brightest sources.

\end{itemize}

A list of candidate sources was constructed by merging the $> 4 \sigma$ excess lists from each mosaic, using a merge radius of 0.1 degrees. A source had to be detected by both search methods in order to be included in the candidate list. Manual inspection was performed on each map to check for the (rare) occasions where {\it SExtractor} fails due to the close proximity of two sources, and any additional sources found were added to the excess list. We also added all previous declared {\em INTEGRAL} detections which were not detected in any of our maps, in order to be able to search for them on different timescales in the later analysis (see section~\ref{bursticity}).

This resulted in a list of 1266 excesses which were passed to the next stage of analysis.

\subsection{Light curve generation and search on all timescales \label{bursticity}}

The main change initiated in this catalog compared to previous ones is intended to address the detection of variable sources. The hard X-ray sky is extremely variable, and this leads to problems in detecting sources when the search is only performed on a limited number of timescales. As an example, a source in outburst in the early mission becomes of lower and lower significance as more and more data is acquired when the source is in quiescence.

These variability issues have led to a number of unfortunate effects:

(1) sources detected in earlier catalogs may drop below the detection threshold after long periods of quiescence.

(2) a number of sources known to be detectable in IBIS have not been included because the source search was not optimised for the particular timescale on which the source was active.

In this catalog, we have performed a systematic search for any source detected in a previous IBIS catalog or declared as a new IBIS detection in the literature prior to April 2008 when the dataset was frozen. This was performed by creating a light curve for each source in the 18-60 keV band on science window timescales, and then scanning a variable-sized time window along each light curve. The window length is varied from 0.5 days ($\sim$10 scw) to the full length of the light curve, and all data points within the time window are included in the analysis. The duration and time interval over which the source significance is maximised is recorded. We define the {\em bursticity} of a source as the ratio of the maximum significance on any timescale, compared to the significance defined for the whole dataset. Thus a bursticity of 1 defines a persistent source, where the inclusion of any data maintains or increases the detection significance. Conversely, a bursticity of greater than 1 implies that the significance of a source can be increased by the omission of some observations from the analysis, presumably when the source was in quiescence. Note that we only use the single time interval when the significance is maximised, we do not combine multiple non-consecutive outbursts which, for some sources, could yield an even higher significance.

The impact of this bursticity analysis is significant. Around 100 sources are recovered that would not have been without this analysis. Furthermore, by defining the time interval over which the significance is maximised for every source, we gain an insight into the variability behaviour of the sources. Finally, by building a mosaic map only for the timescale of maximum significance, we can optimise the chance of source detection and determination of some source parameters - notably the best possible source position (since error radius is inversely proportional to significance).

A few examples can serve to illustrate the effect of the bursticity analysis. The first is IGR J00245+6251, a GRB reported in the third IBIS/ISGRI catalog as an 11.5$\sigma$ detection in revolution 266. The bursticity analysis instead identifies that the source was active on a 0.5 day timescale (this is actually the minimum search time, and is still much longer than the burst itself). By mapping on a more appropriate timescale, the significance is increased to 28.6 sigma and the position error reduced from 2.3$'$ to 1.1$'$. It should be noted that the majority of GRB and a number of other fast (duration $<$ 0.5 days) transient objects with lower fluxes are still not recovered with sufficient significance to be included in this catalog. The second is IGR J17191-2821, a transient discovered during the Galactic Bulge monitoring \citep{Kuulkers2007}. This source was therefore added to the checklist as a previously declared {\em INTEGRAL} detection, despite the fact that it was totally undetectable (below 4$\sigma$) in any of the long-term maps. It would not have been detected using the methods employed for the third catalog. Bursticity analysis, however, confirms its detection at the 8$\sigma$ level during a 1.2 day outburst. These two examples show the efficiency of finding short outbursts. However, there is another class of detection - non-persistent sources of long duration that are too faint to appear at either revolution or whole-archive timescales. Illustrating this is IGR~J13400-6429, a source put forward for further analysis due to a marginal detection ($4.0 <  \sigma < 4.5$) in the whole-dataset maps. Mapping over the optimum 500-day period identified by bursticity analysis provides a clear 7.5$\sigma$ detection, but again this source would not have been found by the methods used for the third catalog.

At the end of this process, the significance of each source in each energy band, and for both whole dataset and `outburst' are known, and these significances are used in a final decision on the acceptance of each source. An indication of the bursticity level, the significance obtained, and a peak flux during the detected outburst are included in the source list (Table~\ref{table:sources}).

\subsection{Source list final filtering}

We have performed a number of steps to minimise the possibility of false catalog entries. These methods are designed to counter both statistical fluctuations in the maps (which we can to some extent assess) and systematic effects present in the maps, which are much harder to quantify. 

First and foremost, each source is manually inspected by a number of people experienced with working with IBIS/ISGRI maps. The inspection covers aspects such as PSF shape, consistency across multiple energy bands, and the significance of the source relative to the {\em local} noise levels in the map. We require a unanimous agreement among many viewers that the excess is a true source, a very conservative approach, but one designed to minimise the false detection rate.

A flux-exposure analysis has been carried out in which each detected flux has been compared to the predicted minimum detectable flux for the exposure in which the detection was made. Sources for which the mean flux is much lower than that which could reasonably be detected in a corresponding timescale may have been boosted by systematic effects, or may just be an outlier in the statistical fluctuations of the maps - in either case, the excess is rejected.

\subsection{Detection Significance thresholds}

\begin{figure}[htbp]
   \includegraphics[width=0.95\linewidth,angle=0,clip]{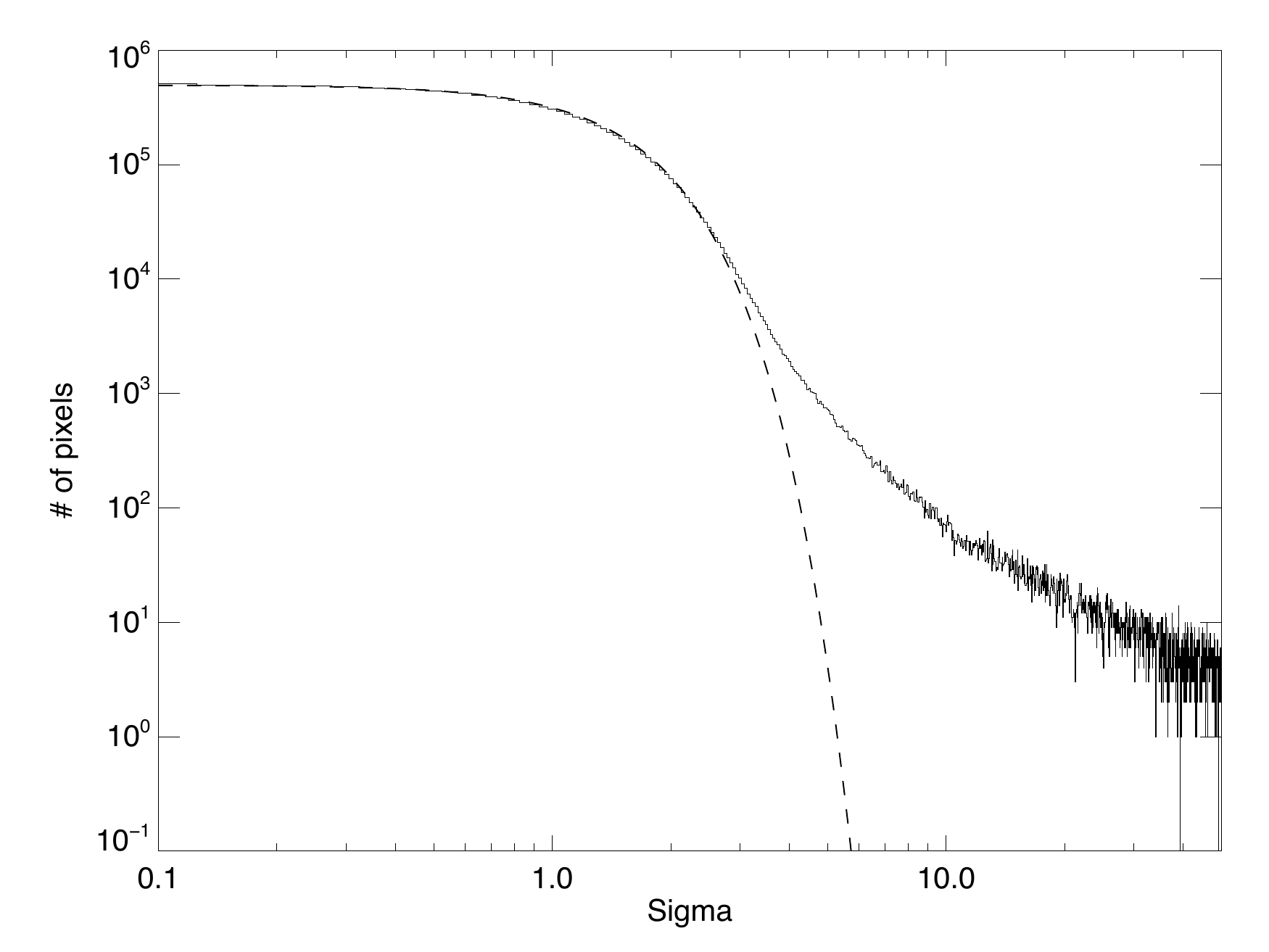}
\caption{Distribution of individual pixel significances found in the
  18--60 keV all-sky mosaic. The solid line represents the data; the
  dashed line represents a a Gaussian fit to the noise distribution.}
\label{fig:c4stats}
\end{figure}

In order to identify an excess in one of the mosaicked images it is necessary to determine the significance level at which the source population dominates over the noise distribution.  To this end we produce a histogram of the individual pixel significance values in each of the mosaics where a source was found. A Gaussian, with mean $\sim$0 and standard deviation $\sim$1, is a found to be a good representation of the noise distribution. This is shown in Figure~\ref{fig:c4stats} for the 18--60 keV all-sky mosaic; at high significances it is clear the data deviates from the noise distribution model.  

Looking at the pixel significance distribution across all mosaics we can confidently conclude that $<<$1\% of the pixels found at significances above 4.8$\sigma$ are produced by the statistical noise distribution. Furthermore, in the 18--60 keV all-sky mosaic, of the pixels found between 4.5--4.8$\sigma$ $<$6\% are from the statistical noise distribution.  However, these limits are based upon the global properties of the mosaics and the maps contain systematic errors which are localised to specific regions. The majority of the systematic noise is produced from the very brightest sources and from very crowded regions. This is dealt with through the visual inspection of each candidate excess in the context of the region of sky in which it has been detected.

\section{Galactic Center Localizations\label{GC}}

The Galactic Center region poses a number of specific problems for the determination of the source population which gives rise to the emission seen by IBIS/ISGRI. A number of the sources there are not fully spatially resolved and are highly variable, and the region is subject to some systematic structures which make the identification of faint sources difficult.

A 50.4$^\prime\times52.8^\prime$ image in the 20--40\,keV band centred at $l=0.12^\circ$, $b=0.18^\circ$ was extracted from each revolution during which the Galactic Center was observed ($\sim140$ in total). The region was optimised to minimise the impact of nearby known bright sources but to allow good assessment of the local background statistics. A core set of three sources, 1E~1743.1$-$2843, SAX~J1747.0$-$2853 and IGR~J17456$-$2901, was used as a starting point, the evidence of their presence being determined from simultaneous, spatially well-separated detection with JEM-X during observations of the Galactic Center and Bulge region performed in revolutions 407--429 (Feb -- April 2007).

\begin{figure}[htbp]
   \includegraphics[width=0.9\columnwidth]{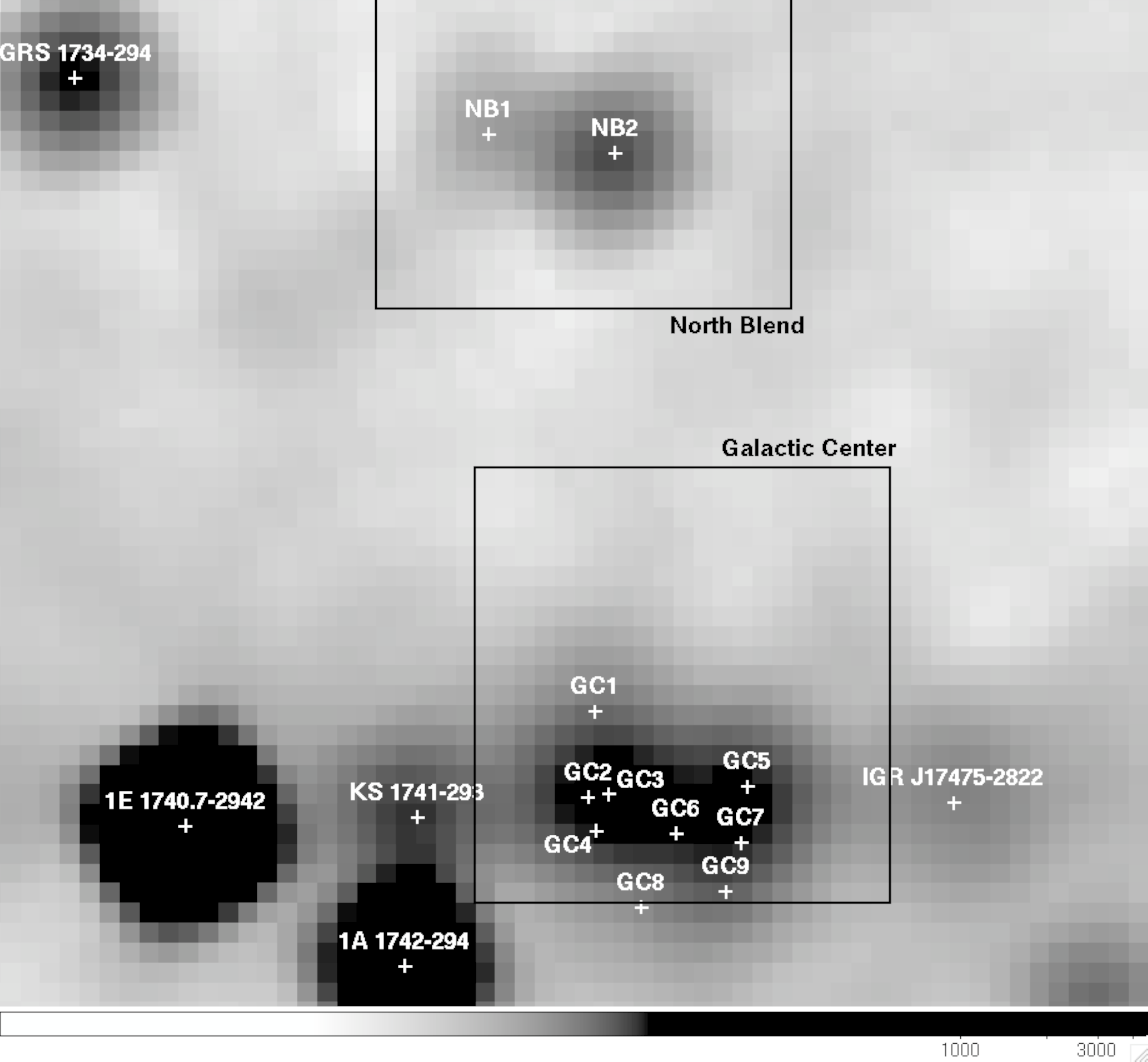}
   \caption{Fitting regions and resulting sources for analysis of Galactic Center and North Blend. \label{fig:GCF1}}
\end{figure}

For each revolution these three sources were fit as two-dimensional Gaussians with their positions fixed to those in \cite{Kuulkers2007}, FWHM fixed to 5 pixels (the PSF of the mosaics described above) and normalisations free to vary. The Gaussian can be taken as an adequate approximation of the true PSF \citep{Gros2003} given that the images are typically constructed from $\sim$50 dithered science windows.  If $\chi^2_\nu >> 1$, and a significant ($>3\sigma$) excess was present in the residuals, a new Gaussian was added with both position and normalisation free to vary. The presence of bright sources centered outside of the fitting region, but still influencing it, was taken into account when necessary. The procedure was repeated for the `North Blend', a region centered at $l=-0.08^\circ$, $b=1.38^\circ$. The two fitting regions are shown in Figure~\ref{fig:GCF1}.

The fit results from all revolutions were merged and a consistent set of nine sources in the Galactic Center and two in the North Blend was generated.  All sources resulting from these fits are identified as B1GCF in the mapcode column of the source list and shown in Table~\ref{tab:GCF}. Apart from the three core sources listed above, we confirm the detection of GRS~1741.9$-$2853 and detect five new variable sources.  The limited angular resolution of IBIS prevents these from being unambiguously associated with previous X-ray sources. The time variability and nature of these sources will be discussed in more detail in a future publication.

\begin{table}[htbp]
\centering
\caption{Sources required in Galactic Centre and North Blend fitting \label{tab:GCF}}
\begin{tabular}{lllll}
ID & Name & R.A. & Dec. & New\\
\hline
NB1 & XTE J1739$-$285 & 264.989 & -28.487 & N \\
NB2 & SLX 1737$-$282 & 265.179 & -28.291 & N\\
GC1 & GRS 1741.9$-$2853 & 266.249 & -28.919 & N\\
GC2 & IGR J17456$-$2901 & 266.410 & -29.021 & N \\
GC3 & IGR J17457$-$2858 & 266.428 & -28.982 & Y\\
GC4 & IGR J17459$-$2902 & 266.485 & -29.043 & Y\\
GC5 & 1E 1743.1$-$2843 & 266.580 & -28.735 & N\\
GC6 & IGR J17463$-$2854 & 266.587 & -28.907 & Y\\
GC7 & IGR J17467$-$2848 & 266.683 & -28.805 & Y\\
GC8 & IGR J17468$-$2902 & 266.690 & -29.045 & Y\\
GC9 & SAX~J1747.0$-$2853 & 266.761 & -28.883 & N\\
\end{tabular}
\end{table}


\section{The Table Data\label{sec:tabledesc}}

The name of the source is given following the convention to quote wherever possible the name declared at the time of the first X-ray detection. The names are given in bold for the $\sim$300 sources added to the catalog since the third catalog.

The astrometric coordinates of the source positions were extracted from the mosaics by the barycentring routines built into {\em SExtractor 2.5}. In almost all cases, the position for a source was extracted from the map yielding the highest source significance. In a few cases, primarily for blended sources, other maps were chosen in order to minimise the interference of other sources. Simultaneous fitting of multiple Gaussian PSFs was used in the most difficult cases - these sources are indicated as blended in the notes accompanying the table. The point source location error of IBIS is highly dependent upon the significance of the source detected \citep{Gros2003}. We use this formulation, combined with the significance of the detection used to locate the source, in order to define an error on the source position. The source localisation errors quoted are for the 90\% confidence limit.

The mean fluxes quoted in the table as $F20-40$ and $F40-100$ are the time-averaged fluxes over the whole dataset derived in two energy bands (20--40 and 40--100 keV). These are provided for compatibility with past catalogs, but we note that their relevance as an {\em average} measure diminishes as the dataset increases and becomes longer than the average time of activity for many of the sources. Therefore, in addition for variable sources, we provide a variability indicator and indicative peak flux in the 20--40 keV band. A flag of Y indicates a bursticity $>1.1$ (ie a 10\% increase in significance can be obtained by selecting a subset of the data. A flag of YY indicates a bursticity of $>4$, indicating a strongly variable source. In both cases, the peak flux is defined as the mean flux during the single period of time for which the significance is maximised.

The type of the source is encoded into up to 4 flags, which are explained in the table footnotes. We have followed the convention of \citep{Liu2007} wherever possible. Identifications, and hence source types, are provided only if considered robust.

The exposure quoted is the total effective exposure on the source after all filtering of the data has been carried out.

The significances quoted are the highest significance in any single map (the map from which the significance is derived is also identified in the table), since this gives the best indication of the robustness of source detection. However, it should be noted therefore that the flux and significance values may derive from different energy bands and/or subsets of the data, and may initially appear contradictory.


\section{Discussion}

We have derived an `unbiased' catalog of 723 sources observed in a systematic analysis of the IBIS/ISGRI Core Programme and public data spanning nearly 5 years of operation. Of these, 684 are secure detections of greater than 4.8$\sigma$, the remainder are detected with between 4.5 and 4.8$\sigma$ but still with a good statistical significance.

\begin{figure}[htbp]
\centering
\includegraphics[width=0.9\columnwidth]{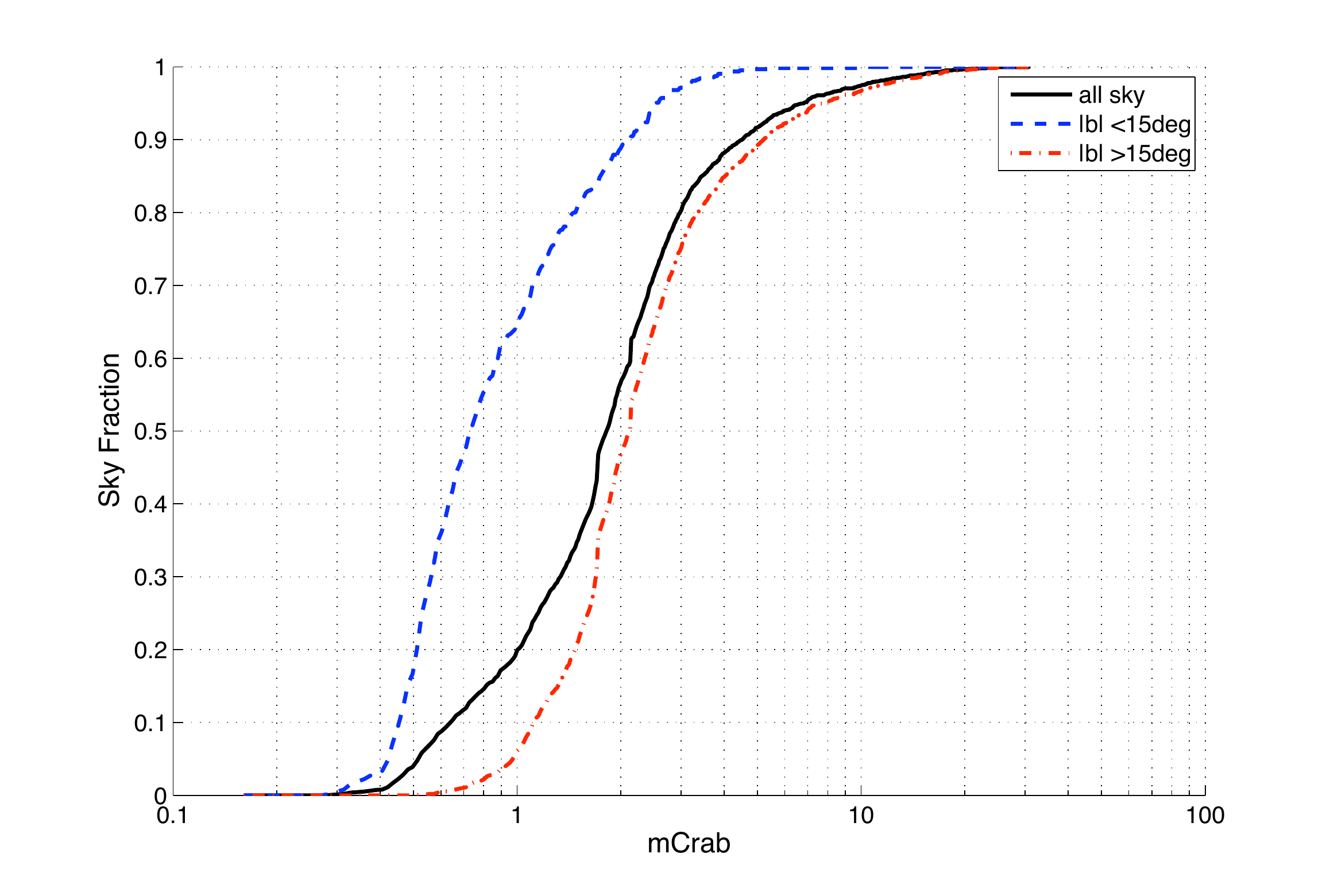}
\caption{Sky fraction as function of minimum detectable flux.
\label{fig:mCrabfrac}}
\end{figure}

We can estimate the minimum detectable flux as a function of the sky position (Figure~\ref{fig:mCrabfrac}) based on the accumulated exposure. The sensitivity of the survey is still strongly biased by the non-uniform exposure. Within the region of the Galactic Plane, $\sim$70\% of the sky is covered to better than 1mCrab sensitivity, while 90\% of the extragalactic sky is now covered at the 5mCrab level.

\begin{figure}[htbp]
\centering
\includegraphics[width=0.9\columnwidth]{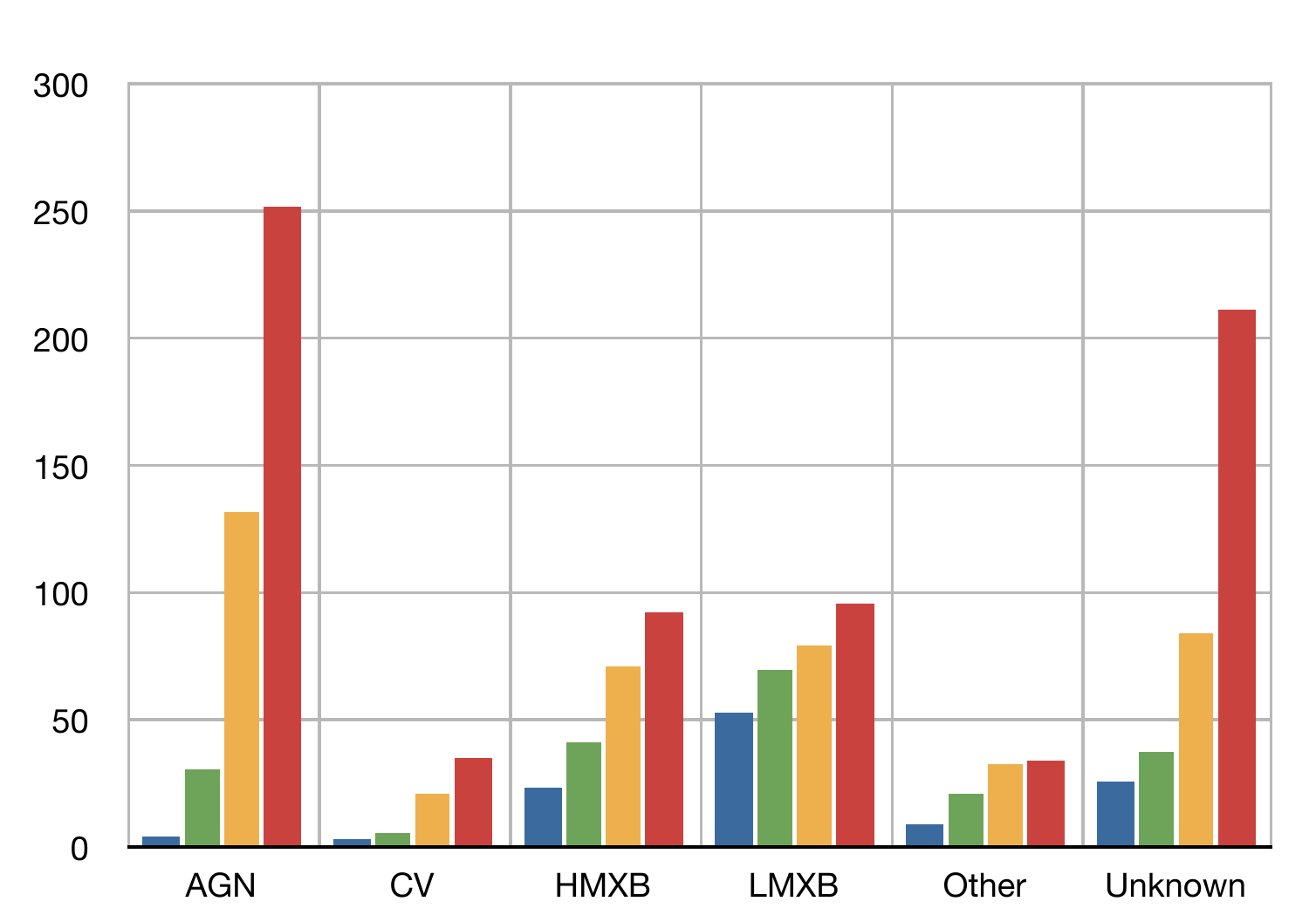}
\caption{Evolution of source type and number through the 4 IBIS/ISGRI catalogs produced to date.
\label{fig:compare1}}
\end{figure}

\begin{figure*}[htbp]
\centering
\includegraphics[width=0.9\columnwidth]{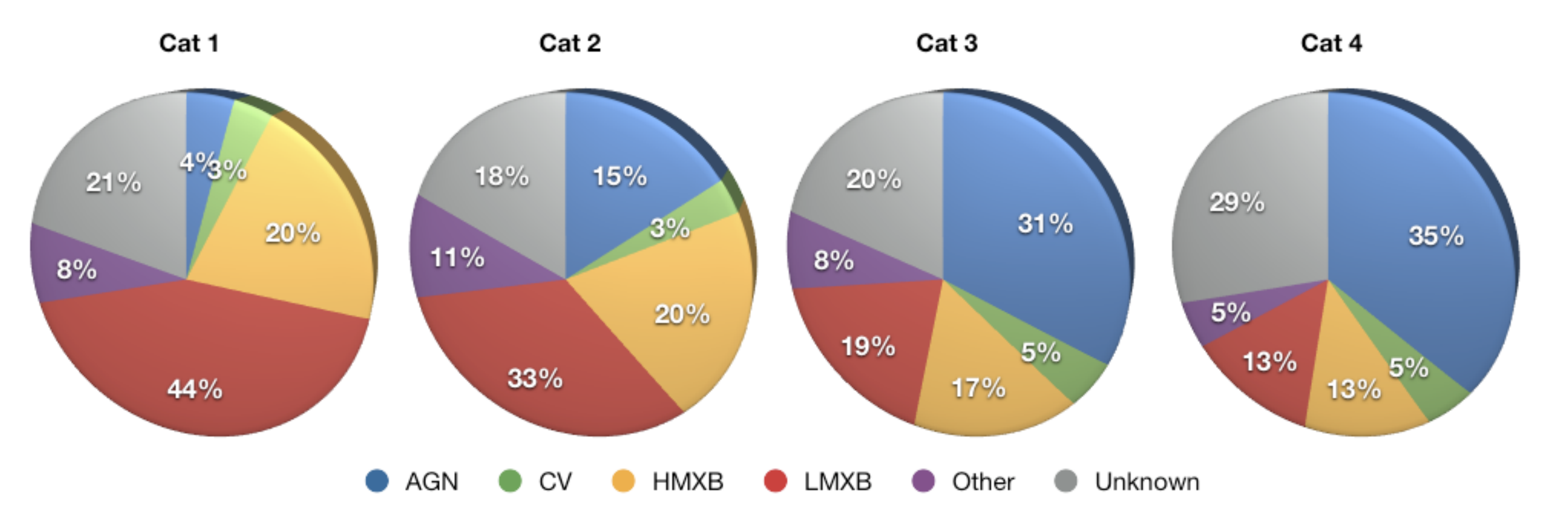}
\caption{Classifications of sources in the 4 IBIS/ISGRI catalogs produced to date.
\label{fig:compare2}}
\end{figure*}

The evolution of the numbers of sources through the 4 IBIS/ISGRI catalogs is shown in Figure~\ref{fig:compare1} and \ref{fig:compare2}. Starting with the first IBIS survey release, we note a continuous increase in the number of extragalactic sources accounting initially for only 4\% of the detected sources in 2005 and now 35\% in the latest source list (see Figure~\ref{fig:compare2})  Experience from previous studies shows that this number will increase further once follow up of the currently unidentified sources can be initiated. It is clear that the changes in the sources dominating the catalogs are strongly linked to the sky coverage. {\em INTEGRAL} spent the first 4 years more on the plane and in particular in the region of the Galactic Bulge while more recently the high latitude sky has been exposed more thoroughly. 

\begin{figure*}[htbp]
\centering
\includegraphics[width=0.9\columnwidth]{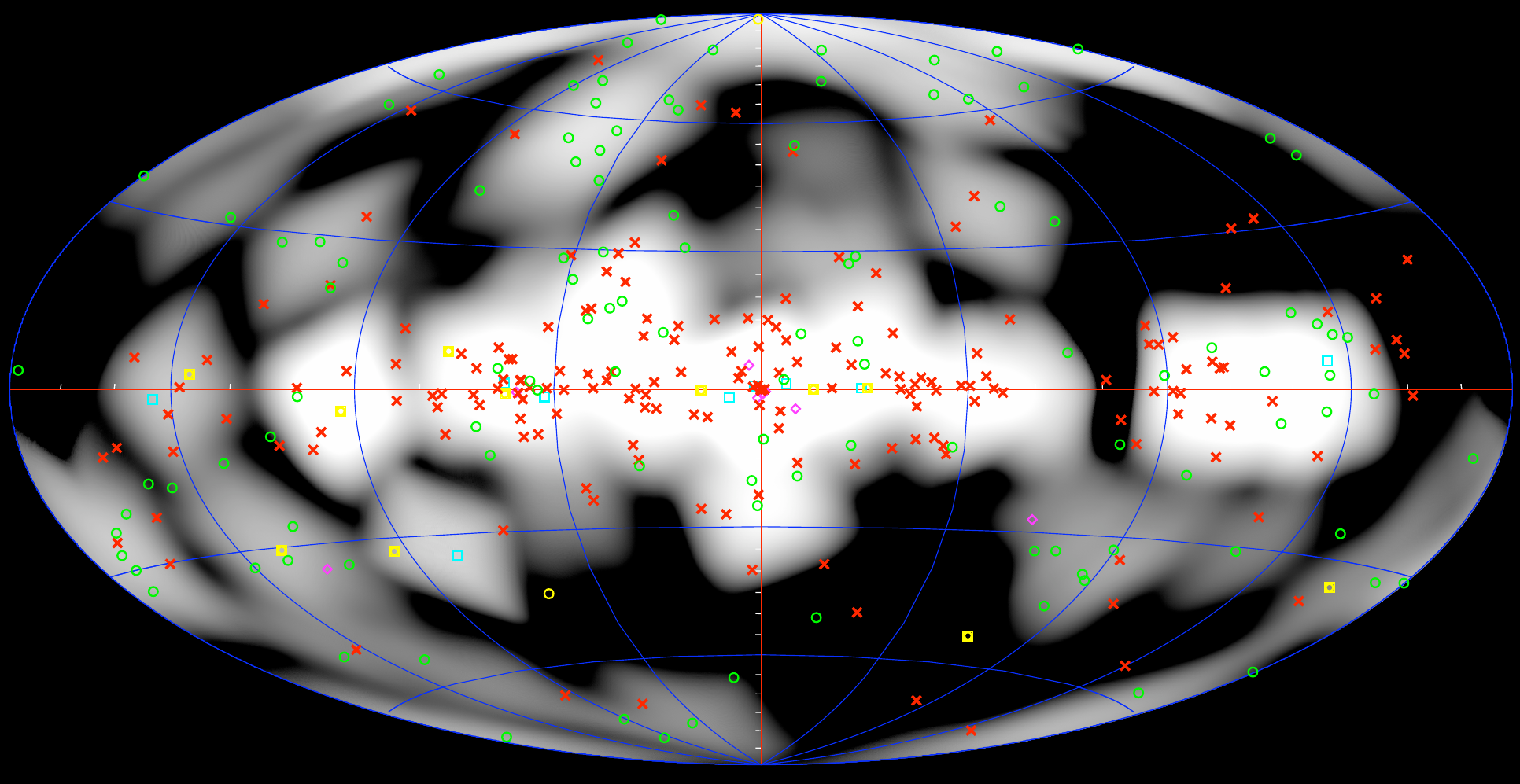}
\caption{Map of incremental exposure since the third catalog, showing the locations of the new sources found. Key: Green circles = AGN; Cyan squares = HMXB; Magenta diamonds = LMXB; Yellow boxes = CVs; Red crosses = Unknown.
\label{fig:newexposources}}
\end{figure*}

There are 331 new sources when compared to the third catalog. Of these, $\sim$120 are associated with extragalactic sources, while only $\sim$25 are associated with known Galactic sources, and the remainder are so far unidentified. This could lead us to conclude that {\em INTEGRAL} is now primarily detecting extragalactic objects and that the survey of the Galactic Plane has reached its limits. However, the sky distribution of new sources (Figure~\ref{fig:newexposources}) shows a rather different picture. When superimposed on the delta exposure (ie the increase in exposure since the third catalog) the new sources can be seen to be following the exposure, and still comprises a very significant Galactic component. We are forced to conclude therefore, that while the extragalactic observations are at a sensitivity limit where IBIS is still re-detecting known objects, the observations near the Galactic Plane have reached a level of depth where previous X-ray observations are no longer always able to provide associations for the new sources. Combined with the variability of the Galactic sources, this is a clear indication that further observations of the Galaxy will continue to uncover new sources, and follow-up of these new sources is of critical importance. However, we should also point out that many of the new sources found in the Galactic Plane by {\em INTEGRAL} have been identified as AGN, so this separation of Galactic and extra-galactic sources is not a straightforward one.

With regards to the `unknown' sources that now constitute nearly 30\% of the source list, one of the main values of this catalog will be to provide hard X-ray sources that will need follow-up at X-ray wavelengths in order to reach a firm identification. To this end, we expect a large fraction of them to be identified in the coming year, as part of an ongoing multi-wavelength campaign. In the third IBIS catalog, 113 sources were not firmly classified. Many of these sources have been followed up at other wavelength starting with an X-ray observation to provide more precise location, allowing for more diagnostic optical or infrared observations. As a result of these observations, 24 previously unidentified sources now have a firm identification and 16 have a tentative but unconfirmed identification. The firm classifications comprise 10 AGNs, 5 CVs, 5 HMXB, 3 LMXB and an XB while tentative classifications are obtained for a further 3 new AGNs, 5 HMXB, 4 LMXB, a CV, a PWN and 2 XBs.

Twenty nine of the sources listed in the third IBIS/ISGRI catalog are not present in this list, for a number of reasons. Three sources (ESO~328$-$IG036, 1RXS J133447.5+371100, MCG$-$02$-$08$-$014) meet all our inspection criteria but are formally below the 4.5$\sigma$ threshold for this fourth catalog. Three of the sources have been removed as part of the reanalysis of the blended regions in the Galactic Center (see section~\ref{GC}). Two sources are no longer detectable because of changes to the dataset (due to use of BTI filtering), and 6 more are rejected as potentially associated with structures around the Galactic Center, LMC or other bright sources. The remaining 16 sources must be considered as likely due to statistical fluctuations in the maps used for the third catalog - indeed 4 of the sources are seen to be associated with low ($<$100ks) exposures in that catalog. The prediction made in the third catalog was for four false sources (representing 1\% of the sample) above the 5--6$\sigma$ cuts, and 10 sources (20\% of the sample) below, whereas the actual numbers are 7 and 9 respectively. Thus we can conclude that the measures taken to quantify false detections from a statistical viewpoint are robust and reasonably accurate. 

We also note the reappearance of one source, IGR~J07506$-$1547, detected in the second IBIS/ISGRI catalog, but not the third. As a result of the bursticity analysis, we are able to confirm the detection of this source that was clearly more active during the early mission phase.

In this catalog, we can state that the detections about $4.8\sigma$ are drawn from an ensemble of maps, all of which show statistical quality that indicates much less than 1\% of the excesses above that level will be false detections. Of the 40 sources below $4.8\sigma$, half are associated with known X-ray emitters, and the estimated $\sim6$\% false detection rate should result in a total number of false detections in this catalog of no more 10, with the vast majority drawn from the sources detected below $4.8\sigma$.

\subsection{Comparison with other hard X-ray catalogs}

It is informative to compare our source list with those coming from other surveys performed in a similar energy range with imaging instruments. 

The first relevant comparison is with the {\em SIGMA/GRANAT} observations performed through 8 years of operational life, providing $\sim$30 Ms of exposure to observe one quarter of the sky, of which 9 Ms were devoted to the Galactic Center region. {\em SIGMA} was characterised by 15$'$ angular resolution and 2--3$'$ accuracy of source location over an energy range of 35 -- 1300 keV. The sensitivity of {\em SIGMA} to the overall sky was about 100 mCrab reaching 8-10 mCrab for the Galactic Center  region. A total of 37 objects were detected above 35 keV, of which 5 were extra-galactic, and 32 galactic, including 8 X-ray novae \citep{Revnivtsev2004a}. The IBIS catalog includes all the {\em SIGMA} extragalactic detections, and all the Galactic ones with the exception of 5 transient neutron star systems (KS~1731$-$26, Tra~X$-$1, GRS~0834$-$43, GRS~1227$-$025, GRO~1744$-$28) and the 8 X-ray novae. The non-detection by IBIS (so far) of these 13 transient systems can be attributed to their long recurrence times between outbursts, together with a low quiescent flux.

Up until the advent of {\em INTEGRAL} and, 2 years later, {\em Swift}, no other imaging instruments were able to improve on the SIGMA results. But now both {\em INTEGRAL} and {\em Swift} are producing surveys of the hard X-ray sky, and a comparison between the most recent results from these missions is informative. The most recent BAT/SWIFT catalog \citep{Cusumano2009} lists 754 sources in the range 14--150 keV derived from 962 detections above 4.8$\sigma$ in at least one of 3 energy bands (14--150, 14--30, 14--70 keV). This is based on 72.7 Ms exposure that is very similar to the IBIS one reached with this fourth catalog. The main difference between the sky as surveyed by the two instruments resides in the ratio of the Galactic and extra-galactic source populations. The IBIS sky in the range 20--100 keV is almost equally shared by Galactic (36\%),  extragalactic (35\%) and unidentified sources (29\%). Conversely, the extra-galactic sources account for 69\% of the BAT list, which contains only 27\% Galactic objects and 4\% of  sources known to be X- or gamma-ray emitters not yet identified.  Within the two lists the most evident difference is the very high number  of blazars from BAT of which IBIS detected only 30\%. This could be explained by both the different exposure/sensitivity, larger FOV and by the flaring activity characterizing these objects. Once parts of the sky recently exposed with {\em INTEGRAL} via the Key Programmes are added to the existing public database, we will be in a better position to fully investigate this difference.   

\begin{figure}[htbp]
\centering
\includegraphics[width=0.8\columnwidth]{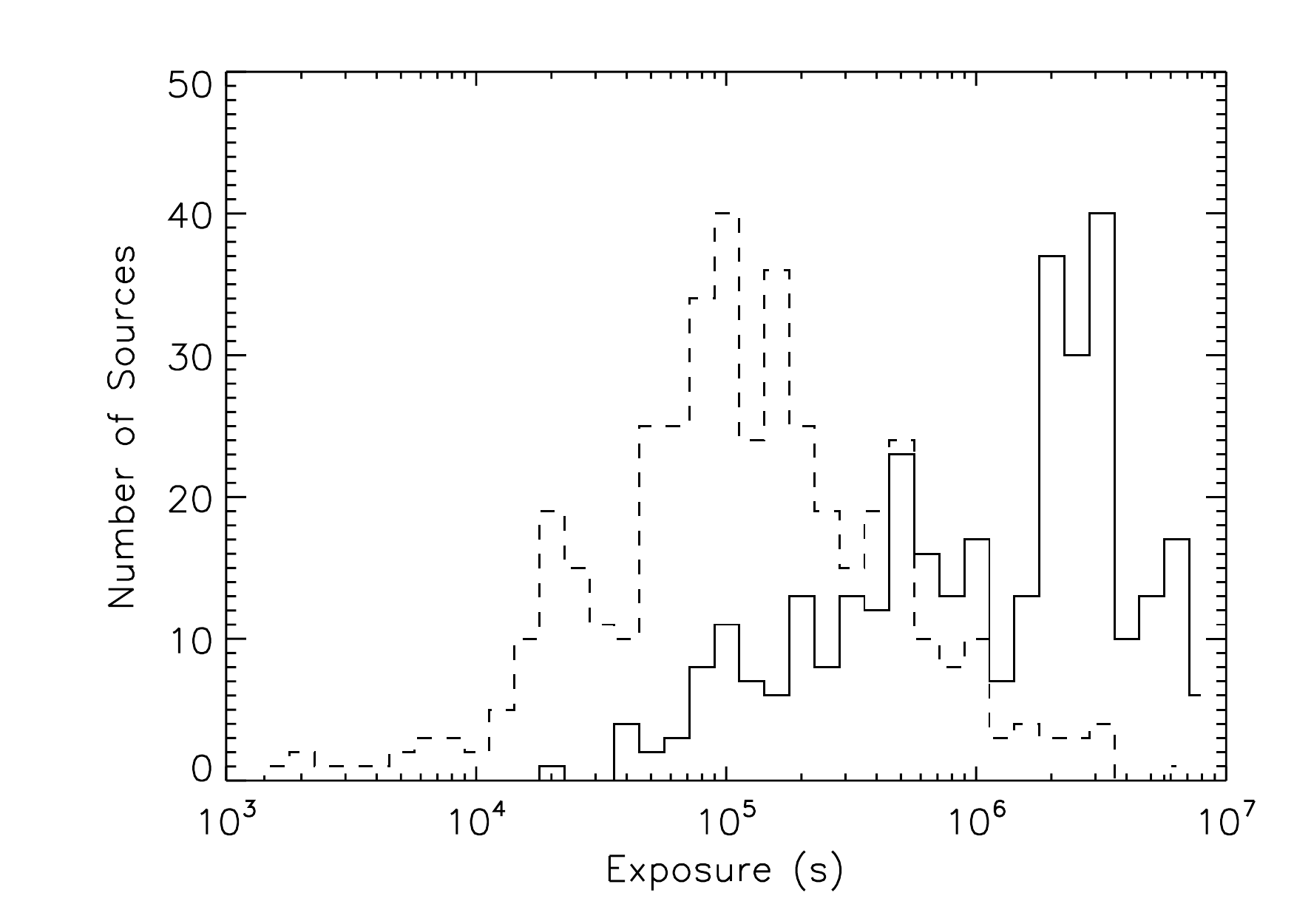}
\caption{IBIS exposure for SWIFT catalogue
sources. The solid line is that for those detected in this catalog whilst
the dashed line is for those not detected. It is clear that
the sources not detected by IBIS have in general a much shorter exposure
time which will account for their not being seen. In the overlap region
around 10$^5$--10$^6$ seconds the lack of detection in IBIS can be ascribed to
source strength and variability reasons.\label{fig:Cat4BatExposure}}
\end{figure}

Cross-correlation of the IBIS and BAT source lists results in 333 correlations within $\sim$400'', the number of false correlations at this level should be around 0. Figure ~\ref{fig:Cat4BatExposure} on the other hand shows the histogram of the exposure for all BAT sources seen in this IBIS catalog (solid line) and the same for all sources not seen in this IBIS catalog (dotted line). Clearly, the great majority of those not seen have a low exposure in IBIS (around 100 below 50ks seconds and another 200 below 200ks). Thus we can conclude that the majority of the differences between the two source lists can be explained by exposure, with any differences at higher exposures likely due to transient sources detected in one or other catalog.

Finally, we note that the current IBIS survey includes all sources reported by the SPI team except one, SPI ~J1720$-$49, for which no further information apart from that the source is variable is available until now \citep{Bouchet2008}. Since the usable SPI sensitivity extends to considerably higher energies than IBIS covers effectively, this implies that there are no sources emitting very hard spectra, or lines above $\sim$200 keV within the SPI sensitive range.

\subsection{Concluding comments}

It is interesting to note the different aim of the {\em INTEGRAL} and {\em Swift} missions that are very clearly demonstrated by the different source populations in the two catalogs. We anticipate that the large difference in the numbers of AGNs with the two lists of sources will be reduced soon once the deep exposures obtained with the {\em INTEGRAL} Key Programmes in AO6 become public and the new AO7 pointings are performed. The current survey shows that IBIS has sufficient sensitivity to detect weak AGNs when these exposures are carried out. Furthermore, the overall picture from the new unidentified sources, accounting for 30\% of our list, indicates the existence of a large galactic population still to be discovered, and we are confident that even in this case the new deep pointings planned for next year (AO7) will result in new source discoveries, possibly new class of objects as in the case of the obscured ones.  The hard X-ray sky requires dedicated observations to solve some of the critical issues currently debated such as the contribution of different types of sources to the X-ray background,  the distribution of intrinsic absorption in sources, and diversity within the same class of objects. {\em Swift} and {\em INTEGRAL}  have been shown to be complementary and have opened new windows of investigations. Moreover, complete and unbiased surveys are of great benefit to studies now being underaken of  the very high energy sky, acting alongside the large soft X-ray database to allow for identification and broad-band analysis of H.E.S.S, MAGIC, VERITAS, AGILE and now FERMI sources. 
 

\acknowledgments

We acknowledge the following funding: Italian Space Agency financial and programmatic support via contracts ASI I/008/07; in UK via STFC grant ST/G004196/1; in France, we thank CNES for support during development of ISGRI and {\em INTEGRAL} data analysis.  This research has made use of: data obtained from the High Energy Astrophysics Science Archive Research Center (HEASARC) provided by NASA's Goddard Space Flight Center; the SIMBAD database operated at CDS, Strasbourg, France; the NASA/IPAC Extragalactic Database (NED) operated by the Jet Propulsion Laboratory, California Institute of Technology, under contract with the National Aeronautics and Space Administration.


\clearpage

\begin{deluxetable}{lccccclccccl} 
\rotate
\tabletypesize{\scriptsize} 
\tablecaption{4th IBIS/ISGRI Catalog \label{table:sources}} 
\tablewidth{0pt} 
\tablehead{ 
\colhead{Name\tablenotemark{a}} & 
\colhead{RA} & 
\colhead{Dec} & 
\colhead{Error\tablenotemark{b}} & 
\colhead{F20-40\tablenotemark{c}} & 
\colhead{F40-100\tablenotemark{c}} & 
\colhead{Type\tablenotemark{d}} & 
\colhead{Vari\tablenotemark{e}} & 
\colhead{Peak flux\tablenotemark{f}} & 
\colhead{Signif\tablenotemark{g}} & 
\colhead{Exposure\tablenotemark{h}} & 
\colhead{MapCode\tablenotemark{i}} 
} 
\startdata 
IGR J00040+7020        &  0.960 & 70.305 & 3.2 &  0.7$\pm$0.1 &  1.0$\pm$0.2 & AGN,Sy2 &  &  &   8.4 & 2613.7& B5 \\ 
{\bf IGR J00158+5605} &  3.961 & 56.092 & 4.9 &  0.5$\pm$0.1 & $<$ 0.3 & AGN? & Y &  1.7$\pm$1.1 &   5.1 & 3030.8& B1 \\ 
IGR J00234+6141        &  5.738 & 61.677 & 2.9 &  0.7$\pm$0.1 & $<$ 0.3 & CV,IP &  &  &   9.2 & 3767.5& B4 \\ 
IGR J00245+6251        &  6.115 & 62.835 & 1.1 & $<$ 0.2 & $<$ 0.3 & GRB & YY & 21.3$\pm$0.9 &  28.6 & 3797.7& burst \\ 
4U 0022+63        &  6.313 & 64.133 & 2.5 &  0.7$\pm$0.1 &  0.6$\pm$0.1 & SNR &  &  &  10.5 & 3729.3& burst \\ 
IGR J00256+6821        &  6.329 & 68.357 & 2.6 &  0.7$\pm$0.1 &  1.0$\pm$0.2 & AGN,Sy2 &  &  &   9.6 & 3174.2& B3 \\ 
V709 Cas        &  7.198 & 59.284 & 0.6 &  4.5$\pm$0.1 &  2.7$\pm$0.1 & CV,IP &  &  &  56.9 & 3562.7& B5 \\ 
IGR J00291+5934        &  7.255 & 59.564 & 0.4 &  2.4$\pm$0.1 &  2.8$\pm$0.1 & LMXB,XP,T & Y & 29.8$\pm$0.3 & 113.3 & 3580.2& burst \\
\enddata
\end{deluxetable}

\end{document}